\documentclass[11pt]{article}
\usepackage{titlesec}
\usepackage{amsmath,esdiff}
\usepackage{amssymb}
\usepackage{authblk}

\usepackage[normalem]{ulem}
\usepackage{bbold}
\usepackage{BOONDOX-cal}
\usepackage[colorlinks=true,linkcolor=blue,citecolor=teal,urlcolor=teal]{hyperref}%
\usepackage{geometry}

\usepackage{mathtools}
\usepackage{subcaption} 
\usepackage{url}

\usepackage{setspace}
\usepackage{cite}
\usepackage{flushend}
\usepackage{mathrsfs}
\usepackage{hyperref}
\usepackage{graphicx}
\usepackage{cancel}

\usepackage{array,esint}
\usepackage[most]{tcolorbox}
\graphicspath{}
\titleformat{\section}{\fontsize{12}{12}\bfseries}{\thesection}{1em}{}
\twocolumn

\begin{document}
\twocolumn[\begin{@twocolumnfalse}
\title{\textbf{Rényi Law Constraints on Gauß-Bonnet Black Hole Merger}}
\author{\textbf{Neeraj Kumar${}^{a ~*}$, Ankur Srivastav${}^{b~\ddagger}$, Phongpichit Channuie${}^{a~ \dagger}$}}
\affil{{${}^{a}$ School of Science, Walailak University}\\
{Nakhon Si Thammarat, 80160, Thailand}\\
{${}^{b}$ Vahrenwalder Str., Hannover-30165, Germany}\\
{}}
\date{}
\maketitle
\begin{abstract}
\noindent In this article, we explore the Rényi law constraints on black hole merger in Gauß-Bonnet (GB) gravity. Specifically, we consider the case of static solutions in five-dimensional (5D) Anti-de-Sitter (AdS) spacetime and study the constraints on merger of two equal mass black holes. We calculate the general Rényi entropy expression and utilize it to study the bounds on the final black hole mass post-merger. We study its variation with the Rényi parameter. We also compare the results with those for black holes in General Relativity (GR). We find that the GB term has a significant impact on the bounds for black hole merger. The bounds for GB gravity become weaker for the zeroth order Rényi entropy and stronger for higher order Rényi entropies in comparison to GR.  
\end{abstract}
\end{@twocolumnfalse}]
\section{Introduction}
\noindent\let\thefootnote\relax\footnote{{}\\
{$*$nkneeraj06@gmail.com}\\
{$\ddagger$ankursrivastavphd@gmail.com}\\
{$\dagger$}phongpichit.ch@mail.wu.ac.th}
\noindent Hawking's area theorem \cite{area1} is a fundamental result in GR which provides constraints on the evolution of black holes. Specifically, in the case of a black hole merger \cite{frans, Shapiro, Abbott}, it provides a bound on the final mass of the resulting black hole and the maximum amount of energy that can be radiated away in the form of gravitational waves. The importance of this theorem lies in the fact that it bounds the final state of a gravitational system without requiring the explicit solution of the highly non-linear and coupled Einstein equations, which are practically impossible to solve analytically in the merger case. The area theorem also paved the way for black hole thermodynamics, where the area of the event horizon is reinterpreted as the entropy of a thermal system \cite{bek}. In this context, Hawking’s area theorem translates into the second law of thermodynamics, also known as the generalized second law \cite{bek1}. Thus, the generalised second law constrains the state space of black holes, without the need to obtain the full dynamical solutions. \\
\noindent Classically, for macroscopic systems with short-range interactions, the second law of thermodynamics provides a single constraint. However, for systems with small degrees of freedom and beyond short-range interactions, the quantum version of the second law allows for a family of constraints. This family of constraints is written in terms of Rényi divergences and has been shown to be the necessary and sufficient conditions for a system to move towards equilibrium \cite{oppen}. Utilizing gauge/gravity duality \cite{mal, witten, duality_guide}, these Rényi divergences were computed for excited CFT at the boundary and were interpreted in the bulk for out-of-equilibrium black hole systems \cite{alice1}. It was shown that the constraints forbid certain transitions that were allowed by the usual second law of thermodynamics. Similar to the second law, these constraints are also termed as Rényi laws quantified by a single parameter called Rényi parameter. The implications of the these Rényi laws have also been explored for AdS black holes without any reference to the gauge/gravity duality \cite{alice2}. It has been shown that for a variety of black hole merger scenarios the zeroth order Rényi law provides the strongest bound on the final mass \cite{alice2}.\ 

\noindent One important feature of applicability of Renyi laws is that the system under consideration should be in a thermally stable equilibrium, such that a canonical ensemble can be defined for it. Thus, the AdS black holes become a suitable choice. In this regard, the study of Renyi constraints for AdS black holes may provide a window to explore strongly correlated quantum systems by utilising the gauge/gravity duality. Specifically, the gauge/gravity duality has been extensively explored as an unconventional route to address non-trivial problems in QCD \cite{duality_guide}, strongly coupled superfluids/superconductors \cite{ank_2019_epjc, ank_2020_epjc, hartnoll, herzog, ank_2021_prd, ank_2023_prd} and the transport properties of holographic metals \cite{seo, donos, ank_2023_epjc}. A fresh look into these problems awaits the use of Rényi constraints as certain progress in this direction has already been made \cite{renAdS0, renAdS1, renAdS2, uga0, uga1, Dong}. \


\noindent Another important direction is to go beyond GR and explore the applicability of these laws. It is interesting to note that there are strong theoretical reasons  to expect GR to be replaced with some more complete quantum theory, from which GR should emerge in low energy limit. There are also debatable observational issues in early universe cosmology, which warrant possible corrections, both classical and quantum, to GR \cite{kiefer01,kiefer02}. In this endeavour, many modifications to GR have been proposed over the years \cite{gravrev}. One of the simplest modification is to consider higher curvature correction terms, which are well motivated as some low energy degrees of freedom of some candidate theory of quantum gravity \cite{string}. In this paper, we shall consider the black hole merger in the GB gravity theory to study constraints due to Rényi divergences for 5D AdS black holes. GB gravity is also the simplest extension to the GR  that respects Lovelock theorem \cite{love}. Black hole solutions along with their thermodynamic properties in GB gravity have been vastly studied in literature \cite{cai, tim, myung, anni, gla, pedro, neeraj1, neeraj2}, though the list is by no means exhaustive. Here, we shall consider Rényi laws applied to the merger of \(5D\) AdS black holes in GB gravity and extract the bounds on the final mass when two equal mass black holes merge. We shall stick to static, uncharged solutions only.\

\noindent This article is organized in the following way. Section (2) reviews the form of Rényi entropies applied to a quantum system and the additional constraints these impose on it. A static black hole solution along with its thermodynamic properties in GB gravity has been reviewed in section (3). In this section, we shall also derive the Rényi entropy formulae for black holes. Next, in section (4), we consider merger of black holes and study the bounds provided by the Rényi laws. Results are discussed in section (5). 

\section{Rényi Entropies and Monotonicity Constraints}
\noindent Here, we shall briefly review a one-parameter generalization of the standard notion of entropy due to von Neumann \cite{information}. This generalization was proposed by Alfréd Rényi in 1961 \cite{renyi} and the family of entropies thus defined are known as Rényi entropies \cite{hijano}. In general, the n-th order Rényi entropy is defined as, 
\begin{equation}
    S_n(\rho) \coloneqq \frac{1}{1-n}\log{\mathrm{Tr}(\rho^n)}
    \label{renyi_ent_form}
\end{equation}
where \(\rho\) is some normalised density matrix representing the state of the system. In limit \(n\rightarrow0\), \(S_n\rightarrow S_0\), is also known as \textit{Hartley entropy/max entropy} \cite{france}. In this work, the parameter \(n\) would be called the \textit{Rényi parameter}. In an analytical continuation to the complex domain, one may show the following result,
\begin{equation}
    \lim_{n\rightarrow 1}S_{n}(\rho)=S(\rho)\coloneq -\mathrm{Tr}(\rho\log{\rho})~,
\end{equation}
which is the familiar von Neumann entropy \cite{alice2}. \
\noindent There is also a closely related notion of Rényi divergences\cite{information1}, which are defined as, 
\begin{equation}
    D_n(\rho|\sigma)\coloneq\frac{1}{n-1}\log{\mathrm{Tr(\rho^n\sigma^{1-n})}}~. 
\end{equation}
 Here, \(\sigma\) is some normalized reference state. These are the generalization of the Kullback-Leubler divergences \cite{KL}, which can be recovered in \(n\rightarrow 1\) limit, i.e.,
\begin{equation}
    \lim_{n\rightarrow 1}D_{n}(\rho|\sigma)=D(\rho|\sigma)\coloneq\mathrm{Tr}(\rho\log{\rho}-\rho\log{\sigma}).
\end{equation}
In general, these divergences would render the distinguishability of two states. \

\noindent A connection between Rényi entropies and Rényi divergences surfaces if the reference state is a maximally mixed state, \(\sigma=\frac{\mathbb{1}}{d}\)~. Here, \(d\) is the dimension of the Hilbert space. Choosing maximally mixed state as a reference state is relevant to a class of systems for which these states have no dynamics (that is, the Hamiltonian does not change the state)\cite{alice1}. These are thermal equilibrium states which correspond to the limit \(\beta\rightarrow 0\), where \(\beta\) the inverse temperature. The relation between the two quantities, when maximally mixed state is the reference state, is given by
\begin{equation}
D_n(\rho|\mathbb{1}/d)=\log{d}-S_n(\rho)~.   
\label{div_ent}
\end{equation}

\noindent The important properties of these measures, entropies and divergences, are related to the dynamics of the system. If an out-of-equilibrium system moves towards a thermal equilibrium state \(\rho_{\beta}\), which is the reference state, the Rényi divergences at two different times, \(t\) and \(t'\), are related by the following inequality, 
\begin{equation}
    D_n(\rho(t)|\rho_{\beta})\geq  D_n(\rho(t')|\rho_{\beta}) ~~~\forall~~~ t \leq t'~.
    \label{div_con}
\end{equation}
This is a family of monotonicity constraints that a system has to follow on its way to equilibrium. If the equilibrium state is a maximally mixed state, then the monotonicity constraints can be written in terms of Rényi entropies, as apparent from eq.(\ref{div_ent}).\ 

\noindent Using eq.(s)(\ref{div_ent}, \ref{div_con}), the monotonicity constraints in terms of Rényi entropies take the form,
\begin{equation}
    S_n(\rho(t))\leq S_n(\rho(t')) ~~~\forall ~~~t \leq t'~.
    \label{renyi_ent}
\end{equation}
These constraints are referred to as Rényi second laws \cite{alice2}. The second law of thermodynamics corresponds to \(n\rightarrow 1\) limit of eq.(\ref{renyi_ent}). These have been extensively studied in the literature in a variety of contexts \cite{ren1, ren2, ren3 ,ren4, ren5, ren6, ren7, ren8}, in this in-exhaustive list.\

\noindent Before moving to discuss the relevant systems to which these Rényi laws will be applicable, we shall list some inequalities of Rényi entropies associated with the Rényi parameter \(n\) \cite{ineq}. If Rényi entropies are considered as a function of \(n\), then they satisfy the following set of inequalities,
\begin{eqnarray}\nonumber
    S_n(\rho)&\geq&  0\\\nonumber
    \partial_nS_n(\rho)&\leq& 0\\\nonumber
    \partial_n(\frac{n-1}{n}S_n(\rho))&\geq&0\\\nonumber
    \partial_n((n-1)S_n(\rho))&\geq&0\\
    \partial_n^2((n-1)S_n(\rho))&\leq&0~.
    \label{en_in}
\end{eqnarray}
One important property of Rényi entropies which distinguishes them from the von Neumann entropy is its dependence on ensembles. It has been explicitly shown in \cite{alice2} that Rényi entropies carry the information of fluctuations and hence the ensemble equivalence in thermodynamic limit does not hold. Based on this property, following considerations are important while applying Rényi laws to a thermodynamic system. For explicit details on following points, please refer to \cite{alice2}.
\begin{enumerate}
    \item In a microcanonical ensemble, Rényi laws reduce to ordinary second law and renders no new constraints (information).
    \item For homogeneous systems, Rényi laws again provide no new information. In other words, for systems which follow the scaling laws of the form \(S(\lambda M, \lambda Q)=\lambda^bS(M, Q)\), with some scaling parameter \(\lambda\) and exponent \(b\), the Rényi entropy formulae reduces to \(S_n=g(n)S\). Thus, no new constraints arise when applied to a homogeneous system. 
    \item Systems in canonical and grand canonical ensemble will have new constraints, however, the condition of a reference state to be maximally mixed state at equilibrium demands the system to be at a stable equilibrium in limit \(\beta\rightarrow 0\).
    \item For systems in unstable equilibrium, the inequalities mentioned in eq.(\ref{en_in}) fail to hold. Thus, along with \(\beta\rightarrow 0\), the system should also be in stable equilibrium in the range of applicability.  
\end{enumerate}
Particularly, with the above conditions in mind, the application of Rényi laws for black holes become important in AdS spacetime as black holes are stable and it is possible to define canonical ensemble for a certain range of parameters. Constraints on black hole merger, in GR for AdS black holes, under different scenarios were considered in \cite{alice2}. Here, we are interested in the application of Rényi laws for AdS black holes in the GB gravity. In the next section, we shall be briefly reviewing the black hole solutions and their thermodynamic properties in GB gravity. We shall then show that they fulfill the above mentioned requirements for the application of the Rényi laws. 

\section{Gauß-Bonnet Gravity and a Static-AdS Black Hole Solution}
One of the simplest extension to GR is the inclusion of the higher curvature terms. A non-trivial one-parameter theory that considers such higher curvature terms, which preserve the diﬀeomorphism invariance and still lead to an equation of motion containing no more than second order time derivatives, is the GB gravity \cite{string, sarkar01}. This special combination of higher curvature terms, also called the GB term, is topological in $D=4$ spacetime dimensions while contributes non-trivially to the equation of motion in $D>4$. In general, an extension to GR with higher curvature terms may be viewed as a low energy effective theory of some UV complete quantum theory of gravity, which is still elusive till date.\

\noindent The gravitational action, with the GB term in a general \(D\) spacetime dimensions with a cosmological constant, may be given as,
\begin{equation}
  S=\frac{1}{16\pi}\int d^Dx\sqrt{-g}(R-2\Lambda+\alpha \mathcal{L_{GB}})
\end{equation}
where \(\mathcal{L_{GB}}=R^2-4R_{\mu\nu}R^{\mu\nu}+R_{\mu\nu\gamma\delta}R^{\mu\nu\gamma\delta}\) is the GB term, \(\Lambda\) is the cosmological constant and \(\alpha\) is the GB parameter. We shall be restricting the values of the GB parameter to be positive. The reason for such a restriction originates in the string theory, which is a candidate unification \cite{string, kiefer01}. \

\noindent A static and spherically symmetric black hole solution for AdS spacetime, \(\Lambda\) $< 0$, in GB theory for $D>4$ can be given in the following form \cite{string}, 
\begin{equation}
    ds^2=-f(r)dt^2+\frac{dr^2}{f(r)}+r^2d\Omega_{D-2}^2
\end{equation}
with the lapse function,
\begin{equation}
    f(r)=1+\frac{r^2}{2\alpha'}\left(1\pm\sqrt{1-\frac{4\alpha'}{l^2}+\frac{4\alpha'm}{r^{D-1}}}\right)~.
    \label{lapse}
\end{equation}
Here, various parameters have their usual meaning, e.g. \(\alpha'\) is related to the GB parameter as \(\alpha'=(D-3)(D-4)\alpha\) and \(m\) is linked to the ADM mass (M) via \(M=\frac{(D-2)\omega_{D-2}m}{16\pi}\) with \(\omega_{D-2}\) being the volume of the unit sphere in \(D-2\) space. The sign, \(`\pm'\) indicates two branches of black hole solutions. However, it was shown in \cite{string} that the black hole solution with the positive sign is unstable. Hence, in this paper, we shall consider the stable branch of solutions and discuss its thermodynamic properties in the next sub-section. 
\subsection{Thermodynamic Properties and Rényi Entropies}
In this section, we shall briefly review the thermodynamic properties of AdS black holes in GB gravity and show that they satisfy relevant conditions related to Rényi laws mentioned before. \

\vspace{.5cm}

\noindent\textbf{Thermodynamic properties of black holes in GB gravity:} The ADM mass of the black hole can be written in terms of the horizon radius, \(r_+\) using equation \(f(r_+)=0\). In our case, it is given as 
\begin{equation}\label{mass}
    M=\frac{(D-2)\omega_{D-2}}{16\pi}\left(\frac{r_+^{D-1}}{l^2}+r_+^{D-3}+\alpha'r_+^{D-5}\right)~.
\end{equation}
Also the Hawking temperature of the black hole can be  obtained from 
\begin{equation}
    T\equiv \frac{1}{4\pi}\frac{\partial f}{\partial r}\Bigg|_{r=r_+}~.
\end{equation}
Using eq.(\ref{lapse}), we get,
\begin{equation}
    T=\frac{(D-1)r_+^3}{4\pi l^2(r_+^2+2\alpha')}+\frac{(D-3)r_+}{4\pi(r_+^2+2\alpha')}+\frac{(D-5)\alpha'}{4\pi r_+(r_+^2+2\alpha')}~.
\end{equation}
We shall now specialize to black holes in \(D=5\) dimensions. This choice is made due to following reasons. First, the GB gravity in $5D$ appears as a first non-trivial higher curvature extension to GR because such an extension in $4D$ becomes topological and does not add anything dynamically. Second, with the advent of the gauge/gravity duality, a $5D$ gravity theory in AdS spacetime is often found to be dual to some $4D$ quantum field theory. From this perspective, investigations of $5D$ GB gravity might be relevant to some real quantum systems. \

\begin{figure}
    \centering
    \includegraphics[width=1\linewidth]{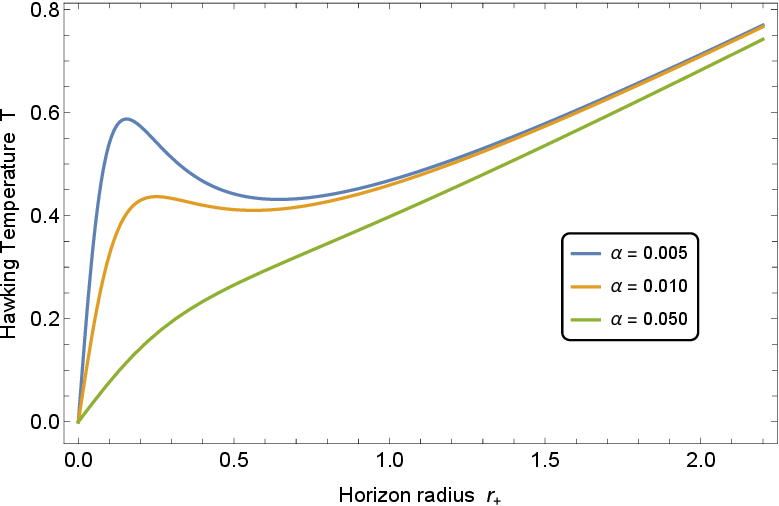}
    \caption{Hawking temperature vs horizon radius for \(5D\) AdS black holes in GB gravity for dimensionless parameter, \(\alpha/l^2=(0.005, 0.01, 0.05)\), and, we set \(l=1\). }
    \label{fig:1}
\end{figure}

\noindent Explicitly, the Hawking temperature in $5D$ takes the form,  
\begin{eqnarray}
T =\frac{r_+^3}{\pi l^2(r_+^2+4\alpha)}+\frac{r_+}{2\pi (r_+^2+4\alpha)}~.
     \label{t5}
\end{eqnarray}
Fig.(\ref{fig:1}) depicts Hawking temperature plotted against the horizon radius, \(r_+\), for different values of the GB parameter.
It is clear from the plot that the black hole is stable in large and small mass range, where temperature gradient is positive. Along with that there is an intermediate range where these are unstable. Our focus in this analysis will be on large mass stable black holes in canonical ensemble. \ 

\noindent Another important modification due to GB term appears in the black hole entropy formula, which gets corrections to the standard area law. Using the first law of black hole thermodynamics, (\(dM=TdS\)), we can calculate the entropy of the black hole as, 
\begin{equation}
\begin{split}
  S  &= \int_0^{r_+} T^{-1}\left(\frac{\partial M}{\partial r_+}\right)dr_+\\
   &= \frac{\omega_{D-2}}{4}r_+^{D-2}\left[1+\frac{2(D-2)(D-3)\alpha}{r_+^2}\right]~. 
   \end{split}
\end{equation}
In $5D$, the above expression reduces to 
\begin{equation}\label{ent}
    S=\frac{\pi^2r_+^3}{2}\left(1+\frac{12\alpha}{r_+^2}\right)~.
\end{equation}
 The black holes are in canonical ensemble, and remains stable in the limit, \(\beta\rightarrow 0\). From eq.(\ref{ent}), it is also clear that the system is not homogeneous. Thus, all four conditions, related to Rényi laws, mentioned in the previous section, are satisfied for the large mass range. \
 
 \noindent This concludes the standard thermodynamic analysis of black holes in GB gravity. Next, we shall focus on calculating the Rényi entropy formula for these black holes. \

\vspace{.5cm}

\noindent\textbf{Rényi Entropies:}
We shall exploit the relation between Rényi entropy formula and free energy as discussed in \cite{free_ent, samolin, alice2} in order to calculate its expression for the black hole under consideration. First, we shall write the free energy for a thermodynamic system in a canonical ensemble. In terms of partition function, it is given by the following expression,
\begin{equation}
    F(\beta)=-\frac{1}{\beta} \ln{(Z(\beta))}
    \label{fr1}
\end{equation}
where \(Z(\beta)=\mathrm{Tr~{e^{-\beta H}}}\) is the canonical partition function. Also, $\beta$ is the inverse temperature i.e. $\beta = \frac{1}{T}$ in units where the Boltzmann constant, $k_B = 1$.\

\noindent Using the expression of thermal density matrix, \(\rho=\frac{e^{-\beta H}}{Z}\) in the Rényi entropy formula given in eq.(\ref{renyi_ent_form}), one gets
\begin{equation}
\begin{split}
    S_n&=\frac{1}{1-n}\left[\ln{(Z(n\beta))}-n\ln{(Z(\beta))}\right]
    \end{split}
\end{equation}
Using eq.(\ref{fr1}), the above expression in terms of free energy becomes
\begin{equation}
    S_n=\frac{n\beta}{1-n}\left[F(\beta)-F(n\beta)\right]~.
\end{equation}
Also, using the relation between free energy and entropy from standard thermodynamics
\begin{equation}
    S(\beta)=\beta^2\frac{\partial F}{\partial \beta}~,
\end{equation}
the Rényi entropy expression can be expressed as \cite{alice2} 
\begin{eqnarray}\label{renent}
    S_n=\frac{n\beta}{1-n}\int_{n\beta}^{\beta}S(\beta')/\beta'^2d\beta'~.
\end{eqnarray}
Now, in order to calculate the Rényi entropies for the black hole under consideration, we shall invert the Hawking temperature in eq.(\ref{t5}) and express it as
\begin{equation}
    \beta=\frac{2\pi l^2(r_+^2+4\alpha)}{2r_+^3+r_+l^2}~.
\end{equation}
This equation can be rewritten as the following cubic equation in $r_+$, 
\begin{equation}
    2\beta r_+^3-2\pi l^2r_+^2+\beta l^2r_+-8\pi l^2\alpha=0~.
    \label{cubic_hor}
\end{equation}
Without any loss of generality, we shall fix the AdS radius, $l$, to unity for further calculations. This shall further simplify eq.(\ref{cubic_hor}) to,
\begin{equation}
     2\beta r_+^3-2\pi r_+^2+\beta r_+-8\pi \alpha=0~.
     \label{eqn}
\end{equation}
In order to have \(r_+\) expressed in terms of \(\beta\), we need to solve the above equation. There are three exact roots to the cubic eq.(\ref{eqn}), however, we are only interested in the largest root as it belongs to the stable phase of the black hole. Our aim in this study is to analyse qualitative effects of the GB parameter, \(\alpha\), on Rényi constraints for black hole merger scenario and hence we shall be restricting the calculations to first order perturbation in \(\alpha\). Note that this means our calculations would be valid for small values of the dimensionless parameter \(\alpha/l^2\), as $l = 1$. Now we consider the following expansion for the horizon radius, \(r_+\), to first order in \(\alpha\),
\begin{equation}
    r_+=r_0+\alpha r_1+\mathcal{O}(\alpha^2)+...
   \label{sol} 
\end{equation}
Substituting this in eq.(\ref{eqn}), we get
\begin{eqnarray}
     2\beta (r_0^3+3\alpha r_0^2r_1)-2\pi (r_0^2+2\alpha r_0r_1)
    +\beta (r_0+\alpha r_1) \nonumber\\ -8\pi \alpha=0~.
    \label{first_order}
\end{eqnarray}
From eq.(\ref{first_order}), we may read off expressions for each order of \(\alpha\) separately as shown below, 
\begin{equation}
    \begin{split}
    \mathcal{O}(\alpha^0):~~&   2\beta r_0^3-2\pi r_0^2+\beta r_0= 0  \\
    \mathcal{O}(\alpha^1):~~&  6\beta r_0^2r_1-4\pi r_0r_1+\beta r_1-8\pi= 0~.
    \end{split}
    \end{equation}
Equation at $\mathcal{O}(\alpha^0)$ provides us with the following relation between horizon radius, in the absence of GB parameter i.e. $r_0$, and the inverse temperature, $\beta$,
\begin{equation}
    r_0=\frac{\pi+\sqrt{\pi^2-2\beta^2}}{2\beta}~.
    \label{r0}
\end{equation}
Here, we have only considered the largest root as it represents the stable phase of the black hole solution. Also, \(r_1\) can be obtained using equation at $\mathcal{O}(\alpha^1)$, 
\begin{equation}
    r_1=\frac{4\pi}{\pi r_0-\beta}~.
    \label{r1}
\end{equation}
Thus, the horizon radius for the black hole in the GB gravity, up to first order in \(\alpha\), takes the following form,
\begin{equation}\label{rsol}
\begin{split}
    r_+&=r_0+\alpha \frac{4\pi}{\pi r_0-\beta}~\\
    &=\frac{\pi+\sqrt{\pi^2-2\beta^2}}{2\beta}+\frac{8\pi \alpha \beta}{\pi(\pi+\sqrt{\pi^2-2\beta^2})-2\beta^2}~.
    \end{split}
\end{equation}

\noindent Now, we shall proceed to calculate the Rényi entropies for the AdS black hole in GB gravity. We start with expanding the expression for the entropy given in eq.(\ref{ent}), again upto first order in \(\alpha\). For convenience, we provide the expression for the entropy here again,  
\begin{eqnarray}
    S=\frac{\pi^2r_+^3}{2}+\alpha (6\pi^2r_+)~.
\end{eqnarray}
Using eq.(\ref{sol}) in the above equation, the entropy to the first order in \(\alpha\), denoted by $\mathcal{S}$, can be written as 
\begin{equation}
\begin{split}
    \mathcal{S}=&\frac{\pi^2r_0^3}{2}+\alpha\left(\frac{3\pi^2r_0^2r_1+12\pi^2r_0}{2}\right)~\\
    =&S^0+\alpha S^1~.
    \end{split}
    \label{en1}
\end{equation}
Here, \(S^0\) is the Bekenstein-Hawking formula representing contribution from GR in 5D and \(S^1\) is the additional contribution due to the GB term. Substituting above form of entropy in the Rényi entropy formula in eq.(\ref{renent}), we get
\begin{equation}
\begin{split}
    S_n&=\frac{n\beta}{1-n}\int_{n\beta}^{\beta}\frac{d\beta'}{\beta'^2} [S^0(\beta')+\alpha S^1(\beta')]\\
    &=S_n^0+\alpha S_n^1~
    \end{split}
\end{equation}
where \(S_n^0\) is due to GR and \(S_n^1\) is the additional contribution due to the GB term. \

\noindent Using eq.(\ref{en1}), the expression of \(S_n^0\) takes the following form, 
\begin{equation}
\begin{split}
    S_n^0=&\frac{n\beta}{1-n}\int_{n\beta}^{\beta}\frac{\pi^2r_0^3}{2G}\frac{1}{\beta'^2}d\beta'\\
      =&\frac{n\beta}{1-n}\left(\frac{\pi}{32G}\right)\left[\frac{(\pi+\sqrt{\pi^2-2\beta^2})^2}{\beta^2}\right.\\
      &\left.- \frac{1}{4}\frac{(\pi+\sqrt{\pi^2-2\beta^2})^4}{\beta^4}\right.\\
      &\left. -\frac{(\pi+\sqrt{\pi^2-2n^2\beta^2})^2}{n^2\beta^2}+ \frac{1}{4}\frac{(\pi+\sqrt{\pi^2-2n^2\beta^2})^4}{n^4\beta^4}\right]~.
\end{split}
\label{ent_zero}
\end{equation}
Where we have used eq.(\ref{r0}) in order to write $r_0$ in terms of $\beta$. It should be noted that in eq.(\ref{ent_zero}) we have recovered the Rényi entropy formula given in \cite{alice2}, in the absence of GB term, for $5D$.\ 

\noindent We have similarly calculated the additional term in the presence of the GB parameter, 
\begin{equation}
    \begin{split}
        S_n^1=&\frac{n\beta}{1-n}\int_{n\beta}^{\beta}\frac{\left(3\pi^2r_0^2r_1+12\pi^2r_0\right)}{2G}\frac{1}{\beta'^2}d\beta'\\
        =&-\frac{n\beta}{1-n}\left(\frac{3\pi}{2G}\right)\\
        &\left[\frac{(\pi+\sqrt{\pi^2-2\beta^2})^2}{\beta^2}-\frac{(\pi+\sqrt{\pi^2-2n^2\beta^2})^2}{n^2\beta^2}\right]~.
    \end{split}
    \label{ent_one}
\end{equation}
Eq.(\ref{ent_one}) gives the first order correction in Rényi entropy in the presence of GB parameter, \(\alpha\). Hence, the complete expression for the Rényi entropy to first order in  \(\alpha\) takes the following simple form,
\begin{equation}
    \begin{split}
        S_n=&\frac{n\beta}{1-n}\left(\frac{\pi}{32G}\right)\left[(1-48\alpha)\left(\frac{(\pi+\sqrt{\pi^2-2\beta^2})^2}{\beta^2}\right.\right.\\
        &\left.\left.-\frac{(\pi+\sqrt{\pi^2-2n^2\beta^2})^2}{n^2\beta^2}\right)\right.\\
        &\left. -\frac{1}{4}\frac{(\pi+\sqrt{\pi^2-2\beta^2})^4}{\beta^4}+\frac{1}{4}\frac{(\pi+\sqrt{\pi^2-2n^2\beta^2})^4}{n^4\beta^4}\right]~.
    \end{split}
    \label{fin_renyi}
\end{equation}
It is interesting to note that the corrections to the Rényi entropy due to the GB gravity are simple but non-trivial. Now, we shall be using eq.(\ref{fin_renyi}) to put constraints on the final mass for a black hole merger scenario in the GB gravity. 

\section{Constraints on Black Hole Merger}
In GR, the second law of black hole thermodynamics provides a bound on the size of the black hole as a result of a merger event. In this section, we shall focus on such bounds due to Rényi laws discussed in previous sections. Note that the mass of a black hole in GB gravity is given as, 
\begin{equation}
    \begin{split}
        M=&\frac{3\pi}{8}\left(r_+^4+r_+^2+2\alpha\right)~.
    \end{split}
\end{equation}
This result can be obtained from eq.(\ref{mass}) by substituting \(D=5\). We have also fixed the AdS radius to unity, as mentioned before. Using eq.(\ref{rsol}), we may find the mass of the black hole, to first order in \(\alpha\), in terms of inverse temperature, \(\beta\). This is given as,  
\begin{equation}
\begin{split}
    M=&\frac{3\pi}{8}\left[\left(\frac{\pi+\sqrt{\pi^2-2\beta^2}}{2\beta}\right)^2 +\left(\frac{\pi+\sqrt{\pi^2-2\beta^2}}{2\beta}\right)^4\right. \\
    &~~~~~~~~~\left.+2\alpha \left(1+ \frac{24\pi \beta}{\pi(\pi+\sqrt{\pi^2-2\beta^2})-2\beta^2} \right)\right]
    \end{split}
    \label{first_order_mass}
\end{equation}
We shall consider head-on merger of two equal sized AdS black holes without spin and charge in $5D$ GB gravity. The Rényi entropy laws then appear as the following constraints,
\begin{equation}
    S_n(M_f)\geq2S_n(M_i)~.
    \label{ent_relation}
\end{equation}
Here, \(M_i\) denotes the initial mass of the coalescing black holes, while \(M_f\) is the mass of the final black hole, that forms after the merger event. Eq.(\ref{ent_relation}) provides additional constraints apart from the standard Hawking area law for different values of the Rényi parameter, \(n\). One should note that for \(n\rightarrow1\), we recover the standard Bekenstein-Hawking entropy constraint. \

\noindent Using eq.(s)(\ref{fin_renyi}, \ref{first_order_mass}, \ref{ent_relation}), we have plotted the final mass, \(M_f\) against the Rényi parameter \(n\) for black holes with different initial mass, \(M_i\).\

\begin{figure}
    \centering
    \includegraphics[width=1\linewidth]{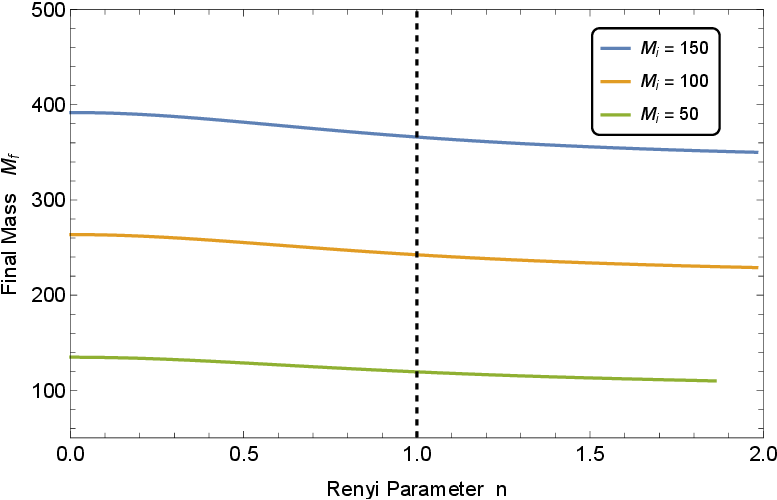}
    \caption{Final black hole mass, \(M_f\), vs Rényi parameter, \(n\), for different initial black hole masses, \( M_i=(50, 100, 150)\) in $5D$ GR.}
    \label{fig:2}
\end{figure}

\noindent Fig.(\ref{fig:2}) shows the variation of the final mass of black holes in $5D$ GR with Rényi parameter (in this case, we have set \(\alpha= 0\)). It is interesting to note here that the qualitative features are similar to the case of black hole merger in $4D$ GR \cite{alice2}. As in the case for the AdS black hole merger in $4D$ GR, here also bounds on the mass of the final black holes become stronger in \(n\rightarrow0\) limit in comparison to bounds imposed by the Bekenstein-Hawking entropy. This means that the zeroth order Rényi entropy formula prohibits the final mass configurations that are allowed by the standard second law of black hole thermodynamics. On the other hand, the bounds due to the higher order Rényi laws are weaker.  However, it should be mentioned that this analysis is valid till \(n=\frac{\pi}{\sqrt{2}\beta}\) as Rényi entropy expression becomes imaginary for larger values of n.\

\begin{figure}[h]
    \centering
    \includegraphics[width=1\linewidth]{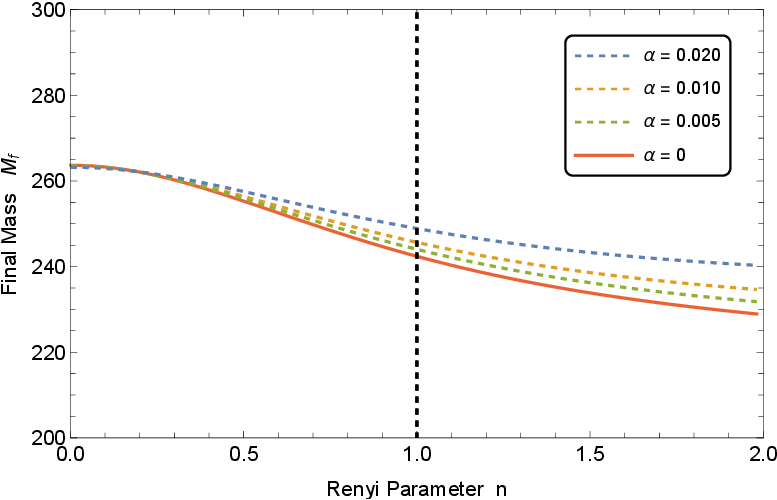}
    \caption{Final black hole mass, \(M_f\), vs Rényi parameter, \(n\), for initial black hole mass, \( M_i=100\) and the GB parameter, \(\alpha=(0, 0.005, 0.01, 0.02)\).}
    \label{fig:3}
\end{figure}

\begin{figure}[h]
    \centering
    \includegraphics[width=1\linewidth]{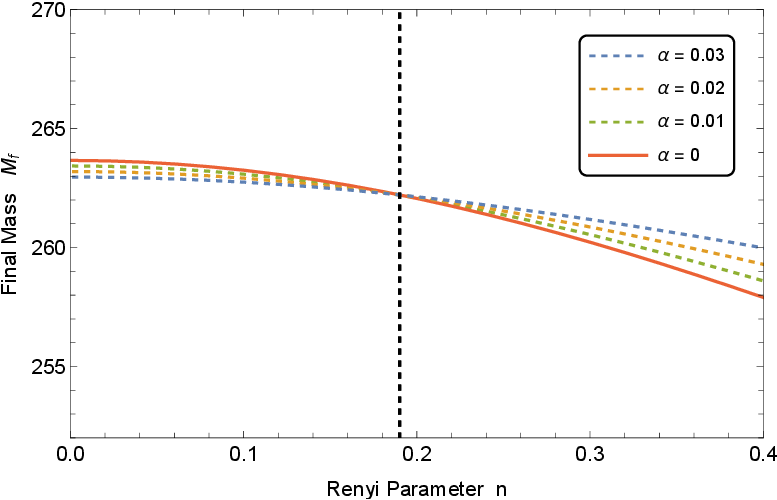}
    \caption{Final black hole mass, \(M_f\), vs Rényi parameter, \(n\), for initial black hole mass, \( M_i=100\) and the GB parameter, \(\alpha=(0, 0.01, 0.02, 0.03)\).}
    \label{fig:4}
\end{figure}

\noindent Next, we shall focus our attention to the impact of the GB parameter on these black hole merger bounds. In Fig(\ref{fig:3}), we have now plotted \(M_f\) against \(n\) for different values of the GB parameter, \(\alpha\). It is clear from the plot that the parameter \(\alpha\) affects the bounds differently for different values of \(n\). A few observations are in order here. Firstly, the first order Rényi law puts stronger bound on the final mass, \(M_f\),  in the presence of the GB parameter, \(\alpha\), in comparison to the Bekenstein-Hawking bound in GR. Second, interestingly the bound imposed by the zeroth order Rényi entropy weaken (see Fig.(\ref{fig:4})). And these impacts, for both cases, become more pronounced with increasing the value of \(\alpha\). Hence, the GB parameter significantly affects the black hole merger bounds.\

\noindent There is another interesting observation regarding these bounds from Rényi entropies with non-integer values of \(n\) between $0$ and $1$. From Fig.(\ref{fig:4}), it is clear that around $n=0.2$, there exist a ``crossover point'' where effects of the GB parameter nullify. Here, the bounds on the final mass of the black holes become equal to that for GR. In other words, the bounds on the final mass at this point are same for the GR and the GB gravity, and it is also independent of the GB parameter. Also for values of \(n\) below the crossover point, the bounds become weaker for the merger of black holes in the GB gravity as the GB parameter is increased. It should be noted though that Fig.(\ref{fig:4}) is plotted for black hole merger with equal initial mass, \(M_i\) = 100.\

\begin{figure*}[ht]
    \centering

    \begin{subfigure}{0.47\linewidth}
        \includegraphics[width=\linewidth]{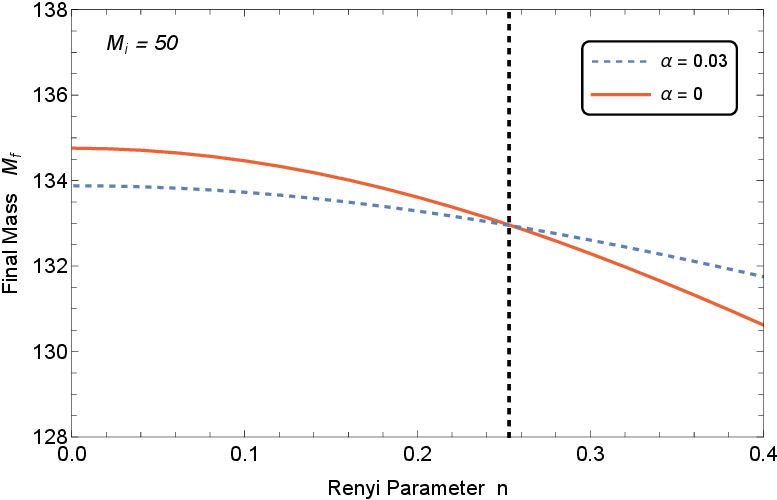}
        \caption{}
    \end{subfigure}\hfill
    \begin{subfigure}{0.47\linewidth}
        \includegraphics[width=\linewidth]{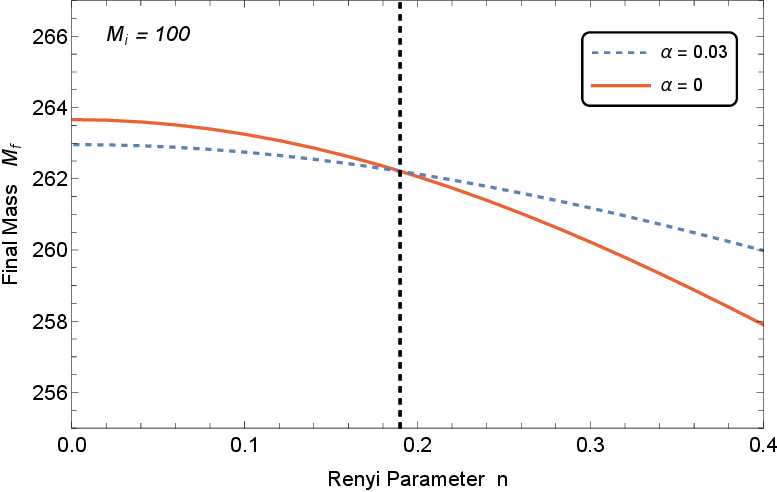}
        \caption{}
        \label{b}
    \end{subfigure}\par

    \begin{subfigure}{0.47\linewidth}
        \includegraphics[width=\linewidth]{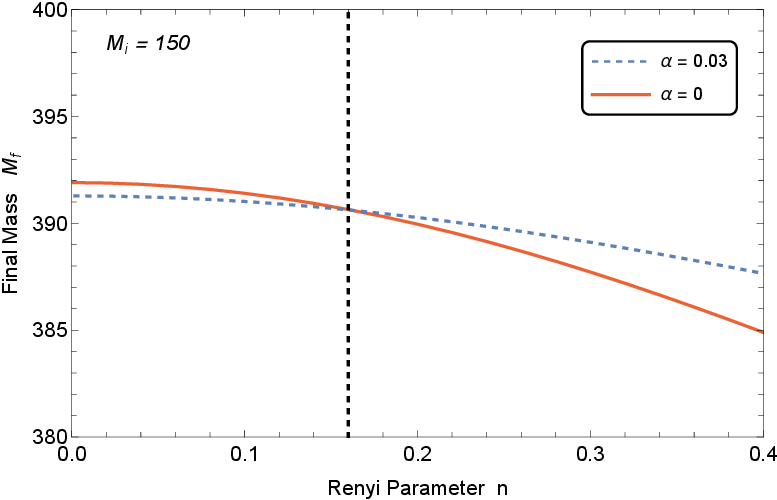}
        \caption{}
        \label{c}
    \end{subfigure}\hfill
    \begin{subfigure}{0.47\linewidth}
        \includegraphics[width=\linewidth]{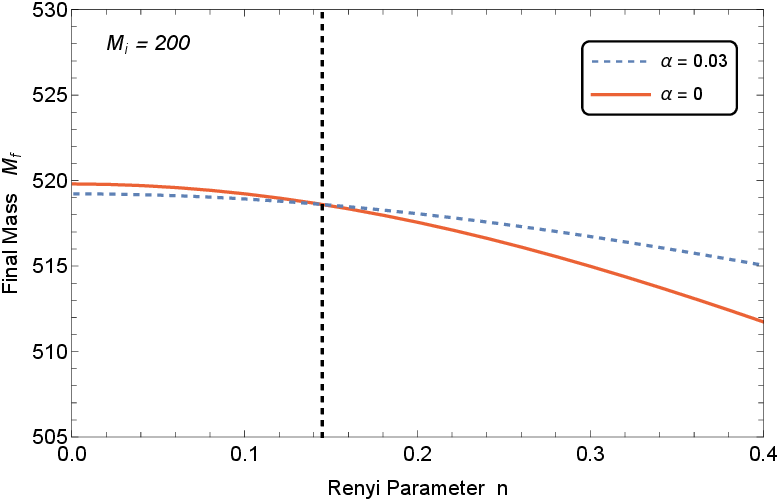}
        \caption{}
        \label{d}
    \end{subfigure}

    \caption{Final black hole mass, \(M_f\), vs Rényi parameter, \(n\), 
    for different initial black hole masses, \( M_i=(50, 100, 150, 200)\) 
    and the GB parameters, \(\alpha=(0, 0.03)\).}
    \label{fig:5}
\end{figure*}

 \noindent We have further plotted the final mass, \(M_f\), with the Rényi parameter \(n\) near the crossover point in Fig.(\ref{fig:5}) for different initial masses, (\(M_i =50, 100, 150, 200)\), for a fixed value of the GB parameter, \(\alpha=0.03\). It is clear from the plots that the crossover point shifts to lower values of \(n\) as the initial mass of the black holes increases. Thus, the zeroth and higher order Renyi entropies in these two gravity theories provide bounds of opposite characteristics about the crossover point. That is, the Rényi bounds in GB gravity, below and above the crossover point become weaker and stronger, respectively, in comparison to GR.  


\section{Discussions}
In GR, the second law of black hole thermodynamics provides a bound on the state space of solutions when gravitational system moves from one state to another. Especially, for the case of black hole merger, the second law provides a bound on the masses of the final black hole solution, which are very informative. Thus, without embarking on the full dynamical analysis of a black hole merger event, one can still comment on the final black hole parameters using the second law.\ 

\noindent A one-parameter extension to the second law of thermodynamics has been posed long ago \cite{renyi} and is know in the literature as Rényi laws. Rényi laws have been shown to be followed by all physical systems on their way to an equilibrium state \cite{free_ent,oppen}. These laws become prominent when dealing with highly correlated systems \cite{oppen}, and they provide additional constraints other than the standard second law of thermodynamics. Rényi laws and the corresponding constraints have been explored in the context of black holes in GR earlier \cite{alice2}. We 
have extended this in the context of the GB black holes in $5D$,  as the GB term only contributes non-trivially in spacetime dimensions greater than four.\ 

\noindent In this work, we have analysed Rényi law constraints in GB gravity for \(5D\) AdS black hole mergers. After listing all the conditions that a black hole system should satisfy so that Rényi laws are applicable, we calculated the Rényi entropy formula for GB black holes upto first order in \(\alpha\). We have then moved to its application and considered the merger scenario where two non-spinning, non-charged black holes of equal masses merge to form a black hole of larger mass. We analysed the bounds on the final mass and studied its variation with the Rényi parameter, \(n\). For black holes in $5D$ GR, We found that the bound is strongest for the zeroth order Rényi entropy and decreases as the value of \(n\) is varied. This result is in synergy with the previous literature in four spacetime dimensions \cite{alice2}. \

\noindent Next, we studied the contrast between both the gravity theories through the black hole merger event and mass bound variation with the Rényi parameter. The bound on merger is weaker for zeroth order Rényi entropy and stronger for higher order Rényi entropies for the GB black holes in comparison to black holes in GR. We have also observed that there is a crossover point for a non-zero value of the GB parameter where bounds from both the theory matches. This crossover point is independent of the value of \(\alpha\) but depends on the choice of the mass of the initial black holes. As highlighted in Fig.(\ref{fig:5}), the crossover point shift to the lower \(n\)-values as we increase the mass of the black holes taking part in the merger process. The impact of the GB term on either side of the crossover point becomes more profound as the value of \(\alpha\)-parameter is increased. Thus, the GB gravity has a significant impact on the final black hole mass bounds. We also wish to highlight the fact that our results are valid in perturbative regime where the GB parameter is small.   \

\noindent   It should be noted that we have considered only stable large mass black holes in \(5D\) GB gravity. However, as shown in Fig.(\ref{fig:1}), there is another branch of small black hole solutions which are also stable, and it is possible to define a canonical ensemble for these solutions as well. Thus, these solutions will also satisfy the conditions that are required to apply Rényi laws. We would like to explore this in future works. It would also be interesting to explore the inequalities associated with Rényi entropies mentioned in eq.(\ref{en_in}) in order to understand the problem of thermal instability of black holes. \

\noindent From holographic perspective, the bounds obtained here would have significant impact on the boundary CFT systems. As shown in Fig.(\ref{fig:1}), there are three different phases of black holes. In this work, we have focused only on large stable black hole phase. However, in order to understand complete phase structure of the boundary CFT, we have to investigate other phases of the black holes as well. In that case, the complete entropy formula would be non-analytic and it should provide insights into the thermodynamics and various phase transitions for the associated boundary CFT. We have left this investigation for the future works.\

\noindent We would also like to emphasize that in this work, we have considered black holes with an additional GB parameter, other than mass. Another important perspective to look into in the future would be to consider black holes with extended phase space (that is, with some charges other than mass) and study the impact of extra charges on the Rényi law constraints. The non-trivial result of weaker bound for the zeroth-order Rényi law for the GB gravity suggests that an extension to black holes with other charges won't be obvious. Specifically, it would be interesting to see the impact for rotating black holes in GR and the GB gravity. In our analysis, we have considered equal mass merger scenario, however, an extension to unequal mass merger seems a trivial exercise, where no new physical insights are expected.\

\section*{Acknowledgement} This research has received funding support from the NSRF via the Program Management Unit for Human Resource and Institutional Development, Research and Innovation grant number B13F680075. AS would like to acknowledge Mrs. Megha Dave for financial support during the work.


\begin{thebibliography}{8}
\bibitem{area1}
S. W. Hawking, 
{Phys. Rev. Lett. 26, 1344 – Published 24 May, 1971}.

\bibitem{frans}
F. Pretorius, 
{Phys. Rev. Lett. 95, 121101 – Published 14 September, 2005}

\bibitem{Shapiro}
S. L. Shapiro, 
{Progress of Theoretical Physics Supplement, Volume 163, May 2006}

\bibitem{Abbott}
B. P. Abbott et al. 
{Phys. Rev. Lett. 116, 061102 – Published 11 February, 2016}

\bibitem{bek}
J. D. Bekenstein, 
{Lett. Nuovo Cimento 4 (1972) 737}.

\bibitem{bek1}
J. D. Bekenstein, 
{Phys. Rev. D 9, 3292 – Published 15 June, 1974}

\bibitem{oppen} 
F. Brandao et al. 
{Proc. Natl. Acad. Sci. U.S.A. 112 (11) 3275-3279}

\bibitem{mal}
J. Maldacena, 
{Adv.Theor.Math.Phys.2:231-252,1998}

\bibitem{witten}
E. Witten, 
{Adv.Theor.Math.Phys.2:253-291,1998}

\bibitem{duality_guide}
M. Natsuume, 
{Lect.Notes Phys. 903 (2015)}

\bibitem{alice1}
A. Bernamonti et al. 
{ J. High Energ. Phys. 2018, 111 (2018)}

\bibitem{alice2}
A. Bernamonti et al. 
{ J. High Energ. Phys. 2024, 177 (2024)}

\bibitem{ank_2023_prd}
A. Srivastav and S. Gangopadhyay, 
{Phys. Rev. D 107, 086005 (2023)}  

\bibitem{ank_2021_prd}
A. Srivastav and S. Gangopadhyay, 
{Phys. Rev. D 104, 126004 (2021)} 

\bibitem{herzog}
C. P. Herzog, 
{J. Phys. A: Math. Theor. 42 343001 (2009)}

\bibitem{hartnoll}
S. A. Hartnoll, 
{Class. Quantum Grav. 26 224002 (2009)} 

\bibitem{ank_2020_epjc}
A. Srivastav et al. 
{Eur. Phys. J. C 80, 219 (2020)} 

\bibitem{ank_2019_epjc}
A. Srivastav and S. Gangopadhyay, 
{Eur. Phys. J. C 79, 340 (2019)} 

\bibitem{donos}
A. Donos and J.P. Gauntlett, 
{J. High Energ. Phys. 2014, 81 (2014)} 

\bibitem{seo}
Y. Seo et al. 
{Phys. Rev. Lett. 118, 036601 (2017)}

\bibitem{ank_2023_epjc}
A. Srivastav et al.
{Eur. Phys. J. C 83, 458 (2023)} 

\bibitem{renAdS0}
A. Belin et al. 
{ J. High Energ. Phys. 2013, 50 (2013)}

\bibitem{renAdS1}
A. Belin et al. 
{ J. High Energ. Phys. 2013, 59 (2013)}

\bibitem{renAdS2}
A. Belin et al. 
{ J. High Energ. Phys. 2015, 59 (2015)}

\bibitem{uga0}
T. Ugajin, 
{ J. High Energ. Phys. 2020, 53 (2020)}

\bibitem{uga1}
T. Ugajin, 
{ J. High Energ. Phys. 2021, 68 (2021)}

\bibitem{Dong}
X. Dong, 
{Nat Commun 7, 12472 (2016)}

\bibitem{kiefer01}
C. Kiefer, 
{International Series of Monographs on Physics (2025)}

\bibitem{kiefer02}
C. Kiefer, 
{arXiv:2302.13047v1 [gr-qc]}

\bibitem{gravrev}
S. Shankaranarayanan, J. P. Johnson, 
{Gen Relativ Gravit 54, 44 (2022)} 

 \bibitem{string}
 D. G. Boulware and S. Deser, 
 {Phys. Rev. Lett. 55, 2656 – Published 9 December, 1985}

\bibitem{love}
D. Lovelock, 
{J. Math. Phys. 12, 498–501 (1971)}

\bibitem{cai}
Rong-Gen Cai, 
{Phys. Rev. D 65, 084014 – Published 25 March, 2002}

\bibitem{tim}
T. Clunan et al. 
{Class. Quantum Grav. 21 3447 (2004)} 

\bibitem{myung}
Y. S. Myung et al.
{ Eur. Phys. J. C 58, 337–346 (2008)} 

\bibitem{anni}
D. Anninos, G. Pastras, 
{JHEP07  030(2009)}

\bibitem{gla}
D. Glavan D and C. Lin, 
{Phys. Rev. Lett. 124, 081301 – Published 26 February, 2020} 


\bibitem{pedro}
P. G S Fernandes et al. 
{Class. Quantum Grav. 39 063001 (2022)}

\bibitem{neeraj1}
N. Kumar et al. 
{Phys. Rev. D 107, 046005 – Published 13 February, 2023}

\bibitem{neeraj2}
N. Kumar, et al. 
{Gen Relativ Gravit 53, 35 (2021)}

\bibitem{information}
M. A. Nielsen, I. L. Chuang, 
{Cambridge: Cambridge University Press (2010)}



\bibitem{renyi} 
A. Rényi, 
{Berkeley Symp. on Math. Statist. and Prob., 1961: 547-561 (1961)}

\bibitem{hijano}
A. May and E. Hijano, 
{J. High Energ. Phys. 2018, 36 (2018)}

\bibitem{france}
 M. Ozawa and N. Javerzat, 
 {EPL 147 11001, (2024)}

 \bibitem{information1}
T. van Erven and P. Harremos, 
{IEEE Transactions on Information Theory, vol. 60, no. 7, pp. 3797-3820, (2014)}

\bibitem{KL}
 S. Kullback and R. A. Leibler, 
 {Ann. Math. Statist. 22(1): 79-86 (March, 1951)}

 \bibitem{ren1}
 M. Müller-Lennert et al. 
 {J. Math. Phys. 1 December 54 (12): 122203 (2013)}

 \bibitem{ren2}
M. M. Wilde et al. 
{ Commun. Math. Phys. 331, 593–622 (2014)}

\bibitem{ren3}
LY. Hung, et al. 
{J. High Energ. Phys. 2011, 47 (2011)}

\bibitem{ren4}
D. A. Galante, et al. 
{ J. High Energ. Phys. 2013, 63 (2013).}

\bibitem{ren5}
D. Petz, 
{Reports on Mathematical Physics, Volume 23, Issue 1, 57-65, (1986)}

\bibitem{ren6}
S. Nojiri et al. 
{Phys. Rev. D 104, 084030 (2021)}

\bibitem{ren7}
S. Nojiri et al. 
{Phys. Rev. D 105, 044042 (2022)} 

\bibitem{ren8}
 E. Elizalde et al. 
 {Universe 2025, 11(2), 60}

\bibitem{ineq}
C. Beck and S. Schogl 
{Cambridge University Press (1993)}
 


 \bibitem{sarkar01}
A. Chatterjee and S. Sarkar, 
{Phys. Rev. Lett. 108, 091301 (2012)} 
 
\bibitem{free_ent}
J. C. Baez 
{Entropy. 2022; 24(5):706}

\bibitem{samolin}
L. Y. Hung et al. 
{ J. High Energ. Phys. 2011, 47 (2011)} 


\end{thebibliography}
\end{document}